\documentclass[12pt]{iopart}
\usepackage{graphicx}
\newcommand\NVO{$\rm Ni_3V_2O_8$}
\newcommand\CVO{$\rm Co_3V_2O_8$}
\newcommand\CUVO{$\beta$-$\rm Cu_3V_2O_8$}
\newcommand\JPhCM[3]{{#3}~{\it J. Phys.: Cond. Matter}~{\bf {#1}}~{#2}}
\newcommand\PRB[3]{{#3}~{\it Phys. Rev. B}~{\bf {#1}}~{#2}}
\newcommand\Kag{Kagom$\acute{e}$}
\begin{document}
\jl{3}

\letter{Single crystals of the anisotropic \Kag\ staircase compounds 
Ni$_3$V$_2$O$_8$ and Co$_3$V$_2$O$_8$}
\author {G. Balakrishnan\footnote[1]{To whom correspondence should be 
addressed, e-mail: G. Balakrishnan@warwick.ac.uk}, 
O.A. Petrenko, M.R. Lees and D.M$\rm ^{c}$K. Paul }
\address{Department of Physics, University of Warwick, Coventry CV4 
7AL, U.K}
\begin{abstract} 
Compounds with a \Kag\ type lattice are known to exhibit magnetic 
frustration. 
Large single crystals of two compounds \NVO\ and \CVO, which are 
variants of a \Kag\ net lattice, have been grown successfully by the 
floating zone technique using an optical image furnace. 
The single crystals are of high quality and exhibit intriguing 
magnetic properties.
\end{abstract}

\maketitle

For a considerable time now, the \Kag\ type lattice has been the 
corner stone of theoretical investigations of highly frustrated 
magnetic systems.
This is because a nearest-neighbour antiferromagnet on a \Kag\ 
lattice (a net of corner sharing triangles) has an infinitely large 
number of classical ground states; the system remains a disordered 
spin-liquid down to zero temperature reflecting this macroscopic 
degeneracy.
From an experimental perspective, however, finding a perfect example 
of a 2D-\Kag\ lattice has proved to be difficult, as even weak 
residual 3D-interactions or magnetic anisotropy can lead to an 
ordered or glassy 
state.

Among the Kagom$\acute{e}$-like systems, the best known and well investigated 
example is $\rm SrGa_4Cr_8O_{19}~$\cite{Ramirez}, while more recently 
a mineral volborthite, $\rm Cu_3V_2O_7(OH)_2 \cdot 
2H_2O$~\cite{volborthite}, $\rm Ba_2Sn_2ZnCr_{7p}Ga_{10-7p}O_{22}$~ 
\cite {chromiumoxide} and different Jarosites~\cite{jarosites}, have 
been advanced as \Kag\ type lattices (for a recent review of the 
chemistry and synthesis of frustrated magnets, including the Kagom$\acute{e}$ 
type, see ref. \cite{Harrison}).

In this letter we describe the growth of single crystals of two 
isostructural compounds \NVO\ and \CVO, by the floating zone 
technique. The compounds $\rm M_3V_2O_8$, where M=Ni and Co, are  new variants 
of the \Kag\ lattice, where the planes that contain the edge sharing 
MO$_6$ octahedra are no longer flat, but buckled to form a staircase 
like structure~\cite{Rogado1,Rogado2}. 
This staircase of magnetic layers separated by the non-magnetic 
VO$_4$ tetrahedra, results in a reduction in the geometric frustration and 
therefore long range magnetic order has been observed in both \NVO\ 
and \CVO ~\cite{Rogado1}. 
While small crystals of both these compounds have been grown earlier, 
from the melt, for crystal structure studies~\cite{Sauer}, large 
single crystals have not been grown so far. 
The magnetic properties of both the Ni and the Co compounds have been 
reported earlier only for polycrystalline powder samples~\cite{Rogado1}.
Neither the exact nature of the observed magnetic transitions nor the 
magnetic ground state of these compounds has yet been investigated. 
In view of the interesting magnetic properties these lattices 
exhibit, it is desirable to carry out investigations on good quality 
single crystals.
Initial characterisation of the grown crystals by X-rays and magnetic 
susceptibility measurements is also presented in this letter.   

Both \CVO\ and \NVO\ adopt orthorhombic crystal structures, space group $Cmca$~\cite{Sauer}. 
Polycrystalline materials were first prepared by the solid state reaction 
of the starting oxides. High purity NiO, CoO and V$_2$O$_5$ were 
taken in stoichiometric ratios and mixed well together.
The mixtures were reacted in air at 800$^\circ$~C for a day.
The Ni compound was further heated at 900$^\circ$~C for a day, while 
the Co compound was heated at 1050$^\circ$~C for two days.
Both powders were ground in between each firing to ensure good 
homogeneity of the mixture.
The polycrystalline powders were checked for the correct phase 
formation and purity using X-ray powder diffraction. 
The powders were then isostatically pressed to produce rods roughly 
7-8~mm in diameter and 70-80~mm long.
The rods were sintered in air at 900$^\circ$~C and 1050$^\circ$~C for 
the Ni and the Co compounds respectively and used for crystal growth. 
Crystal growth was carried out by the floating zone 
technique~\cite{Balakrishnan} in a four mirror image furnace [CSI 
model FZ-T-1000-H-IV-VPS]. 
The crystal growth was carried out at growth speeds ranging from 
0.5~mm/h to 3~mm/h, with the seed and feed rods rotating at 
20-30~rpm.
All the growths were carried out in air at ambient pressure.
A polycrystalline seed rod was used for the first growth and the 
crystal obtained was used as the seed for subsequent growths. 

X-ray Laue photographs were taken to check crystal quality and to 
align the crystals.
Magnetic susceptibility measurements were carried out using an Oxford 
Instruments Vibrating Sample Magnetometer(VSM).

\begin{figure}
\includegraphics[width = 6 cm]{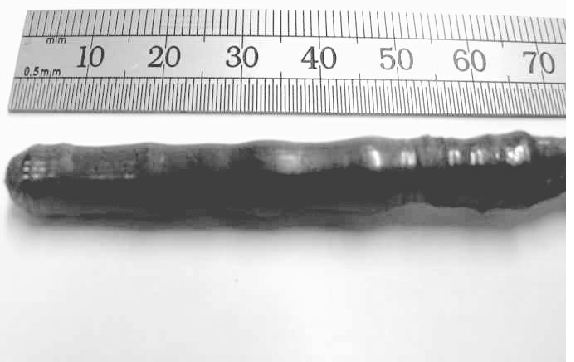}
\includegraphics[width = 7.85 cm]{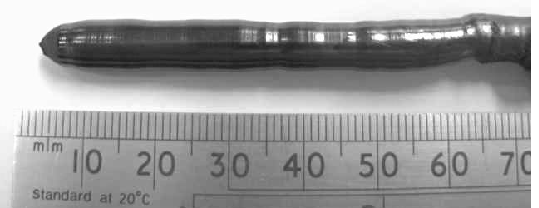}
\caption { Photograph of the as-grown boules of \NVO\ (left) and 
\CVO\ (right) grown by the floating zone technique.} 
\end{figure}
 
Both \NVO\ and \CVO\ have congruent melting points and therefore 
their  crystal growth by the floating zone technique posed no 
problems.
The crystals  developed shiny facets as they grew.
The \NVO\ crystal was dark with a greenish  hue while the \CVO\ 
crystal was black.
Photographs of both the as grown crystal  boules are shown in Fig.~1, 
illustrating the size and volume of the crystals  that can be grown 
by this method. 
The quality of the crystals was checked using X-ray Laue back 
reflection photographs.
A typical photograph obtained from a crystal of \CVO\ is shown in 
Fig.~2.
A few of the crystals grown have also been examined using neutron 
Laue photographs, which have shown that the crystals are suitable for
future inelastic neutron scattering experiments. 

\begin{figure}
\centering
\includegraphics[width = 7 cm]{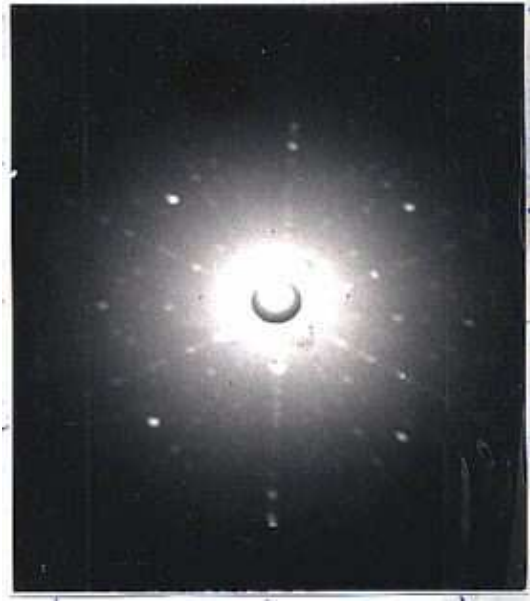}
\caption {X-ray Laue back reflection photograph of a crystal of \CVO\ 
showing the (100) orientation.}
\end{figure}  

Slices were cut from the boule for various measurements and a portion 
of the boule was used for obtaining pieces oriented along the three 
principal crystallographic axes for measurements. Below, we present our initial measurements of 
the magnetic properties of one of the crystals grown, namely, \CVO.  
 The magnetic susceptibility ($\chi$) of a crystal of \CVO\ measured in an 
applied field of 500~Oe and 5~kOe using the VSM is shown in Fig.~3.
In agreement with the results of the magnetic susceptibility 
measurements on powder samples~\cite{Rogado1}, the susceptibility 
is almost Curie-Weiss like at higher temperatures, but small
deviations from $\frac{C}{T-\Theta}$ make the accurate determination 
of the Weiss constant $\Theta$ difficult.

The measurements on the \CVO\ single crystals for $H\parallel a$, 
$H\parallel b$ and $H\parallel c$ reveal the presence of a large 
magnetic anisotropy in this compound.
As can be seen from Fig.~3, the direction of the easy axis 
magnetisation in \CVO\ is along the $c$-axis, while the hard-axis is 
parallel to $b$. Remarkably, the ratio of the magnetic 
susceptibilities for $H\parallel c$ and $H\parallel b$ exceeds a 
factor of 50 at low temperatures.
The susceptibility $\chi_{\parallel c}(T)$ curve shows a sharp increase at a 
temperature around 6.0~K, below which it remains nearly constant 
similar to the powder data \cite {Rogado1}.
It also shows two clear anomalies at $T=8.2$~K and $T=11.2$~K, while 
only one has been seen in powder samples~\cite{Rogado1}.
Both these anomalies $T=8.2$~K and $T=11.2$~K are also evident in the 
corresponding $\chi_{\parallel b}(T)$ curve, which also reaches its 
maximum value at $T=6.0$~K and remains constant at lower 
temperatures. In sharp contrast, the $\chi_{\parallel a}(T)$ curve 
does not show any anomalies at higher temperatures; it reaches its 
maximum at a temperature of $T=6.4$~K and then decreases rapidly by 
about 40\% before becoming temperature independent below $T=5.7$~K.

\begin{figure}
\centering
\includegraphics[width = 9 cm]{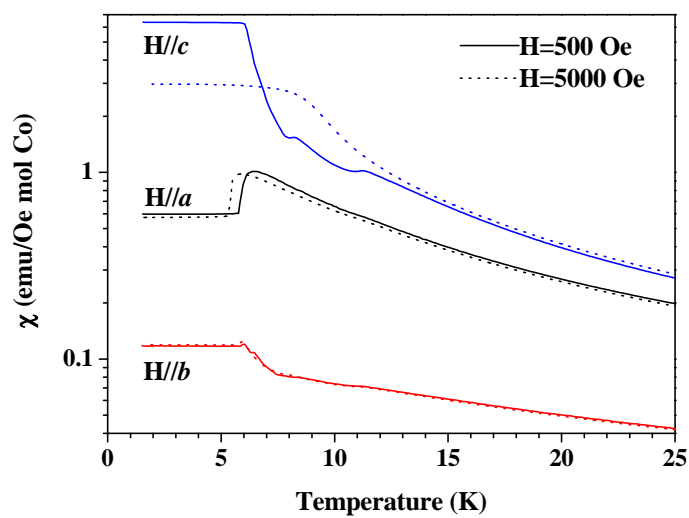}
\caption {Temperature dependence of the magnetic susceptibility of a 
single crystal of \CVO\ measured with the magnetic field applied 
along the three different directions.} 
\end{figure}  

The magnetic susceptibilities for all three directions of applied 
field are quite sensitive to the field values used, with the easy-axis 
susceptibility $\chi_{\parallel c}(T)$  being most affected at the 
higher field (see the dashed lines in Fig.~3).
This observation suggests that \CVO\ may have a complicated $H-T$ phase 
diagram.

Heat capacity measurements reported on powder samples of \NVO\ 
indicate that it exhibits four transitions above 2~K, while the corresponding 
magnetic susceptibility measurement shows no 
indication of two of these transitions (at 6 and 9~K)~\cite{Rogado1}. It will be 
interesting to see if the measurements of magnetic susceptibility on 
single crystal samples provide more detailed information on the 
series of the phase transitions in \NVO . 

In contrast to the multiple phase transitions exhibited by \NVO\ and \CVO, the 
third member of the family \CUVO, is reported to undergo a single 3D 
magnetic transition at $\sim 29$~K with indications of short range or 
low dimensional ordering at higher temperatures~\cite{Rogado2}.
The magnetic properties of \CUVO\ have been investigated only in 
powder form; further progress in the understanding of its magnetic 
properties would undoubtedly benefit from the manufacture of 
single crystal samples.
  
In conclusion, large high quality crystals of the \Kag\ staircase 
lattices \NVO\ and \CVO\ have been grown by the floating zone 
technique.
This method is ideally suited for the crystal growth of this interesting class of 
compounds, especially where large volumes are required for the 
investigation of their magnetic properties by neutron scattering 
experiments.
The magnetic properties measured here on single crystal samples of 
\CVO\ for the first time, show a huge anisotropy.
Specific heat and magnetisation measurements have been carried out on 
these crystals and will be reported separately.
Detailed investigations of the nature of the magnetic transitions in 
these crystals are planned using neutron scattering experiments. 

\ack This work was supported by a grant from the EPSRC, UK (GR/S04024/01).

\References \item

\end{harvard}


\begin{thebibliography}{99}
\bibitem {Ramirez} Ramirez A P, Espinosa G P and Cooper A S \PRB{45}{2505}{1992}
\bibitem {volborthite} Hiroi Z \etal 2001 \JPSJ {\bf 70} 3377
\bibitem {chromiumoxide} Hagemann I S \etal 2001 \PRL {\bf 86} 894 \\
                                               Bono D \etal 
\JPhCM{16}{S817}{2004} \\
                                               Bonnet P \etal 
\JPhCM{16}{S835}{2004}
\bibitem {jarosites} Keren A \etal \PRB{53}{6451}{1996} \\Wills A S 
and  Harrison 1996 {\it A 
J. Chem. Soc. - Faraday Trans.} {\bf  92} 2161
\bibitem{Harrison} Harrison A \JPhCM{16}{S553}{2004}
\bibitem {Rogado1} Rogado N, Lawes G, Huse D A, Ramirez A P and Cava 
R J, 2002 Solid State Commun. {\bf 122} 229 
\bibitem {Rogado2} Rogado N, Haas M K, Lawes G, Huse D A, Ramirez A P 
and Cava R J \JPhCM{15}{907}{2003} 
\bibitem{Sauer} Sauerbrei E E, Faggiani R and Calvo C {1973} Acta. 
Cryst. {\bf B29}  234  
\bibitem {Balakrishnan} Balakrishnan G, Petrenko O A, Lees M R and 
Paul D M$\rm ^c$K \JPhCM{10}{L723}{1998}

\end {thebibliography}
\end {document}